\documentclass{article}

\usepackage{graphicx}

\def\abstract#1{\vskip 7mm 
        \begin{center}{\large Abstract}\par \smallskip
                \begin{minipage}[c]{12cm}
                        \small #1
                \end{minipage}
        \end{center}
}
\def\title#1{\begin{center}{\Large\bf #1}\end{center}}
\def\author#1{\vskip 5mm \begin{center}{#1}\end{center}}
\def\address#1{\begin{center}{\it #1}\end{center}}
\makeatletter
\@ifundefined{lesssim}{}{}
\@ifundefined{gtrsim}{}{}
\def\vereq#1#2{\lower3pt\vbox{\baselineskip1.5pt \lineskip1.5pt
\ialign{$\m@th#1\hfill##\hfil$\crcr#2\crcr\sim\crcr}}}
\makeatother

\begin{document}

\title{%
  PBH and DM from cosmic necklaces
}
\author{%
  Tomohiro Matsuda\footnote{E-mail:matsuda@sit.ac.jp}
}
\address{%
  $^1$Theoretical Physics Group,
  Saitama Institute of Technology,
  Saitama 369-0293, Japan
}

\abstract{Cosmic strings in the brane Universe have
recently gained a great interest.
I think the most interesting story is that future cosmological
observations distinguish them from the conventional cosmic strings.
If the strings are the higher-dimensional objects 
that can (at least initially) move along the compactified space, and 
finally settle down to (quasi-)degenerated vacua in the compactified
space, then kinks should appear on the strings, which interpolate 
between the degenerated vacua.
These kinks look like ``beads'' on the strings, which means that the
strings turn into necklaces. 
Moreover, in the case that the compact manifold is not simply connected,
the string 
loop that winds around a non-trivial circle is stable due to the
topological reason. 
Since the existence of degenerated vacua and a non-trivial 
circle is the common feature of the brane models, it is important to
study  
cosmological constraints on the cosmic necklaces and their stable winding 
states in the brane Universe.}
 
\section{Introduction}
\begin{figure}[h]
\begin{picture}(640,330)(0,0)
\resizebox{16cm}{!}{\includegraphics{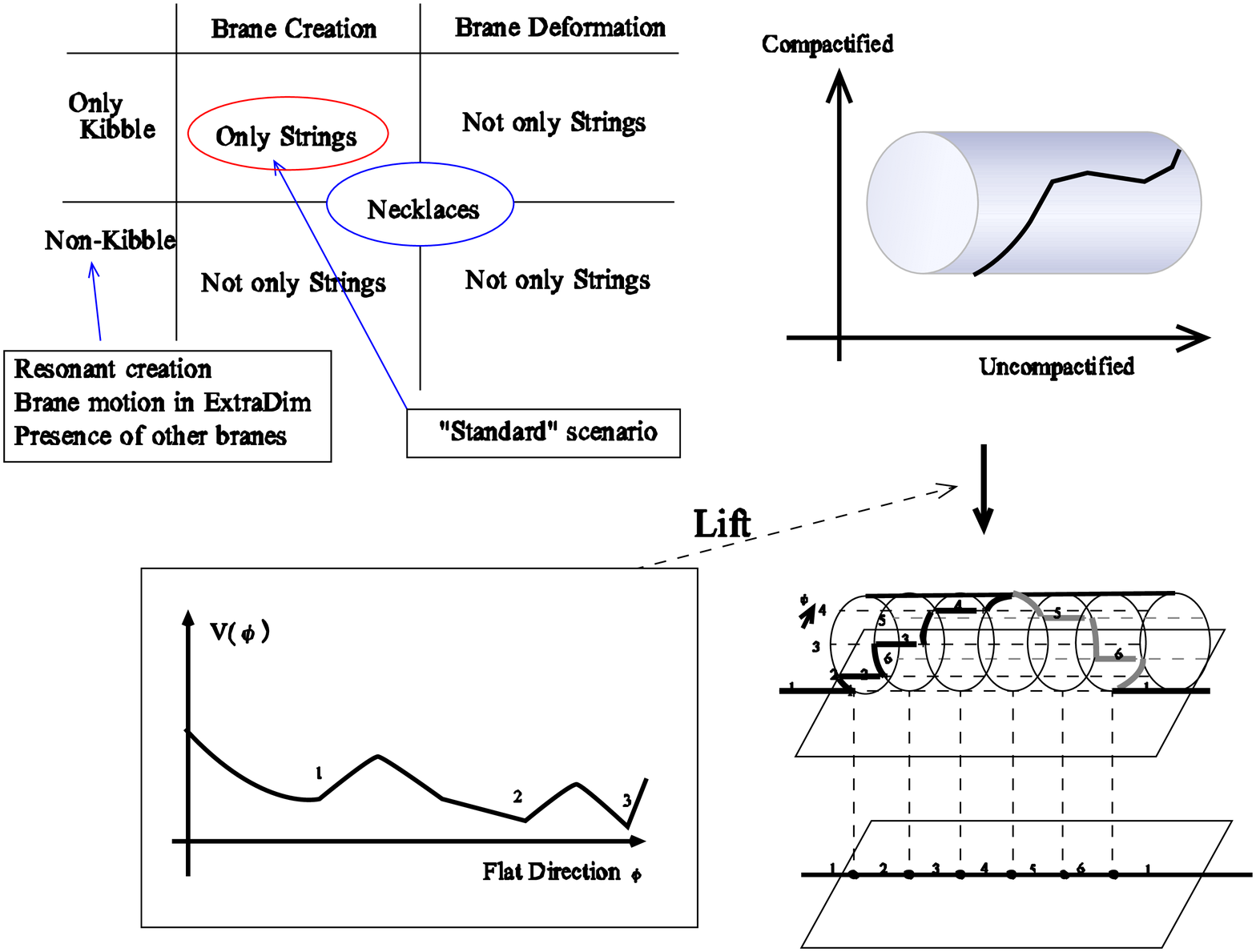}} 
\end{picture}
\caption{We show how necklaces are produced despite the ``standard''
 arguments. } 
\label{fig:fig0}
\end{figure}

In this talk we will explain the cosmological consequences of the
production of Dark Matter(DM) and Primordial Black Hole(PBH) from the
loops of the cosmic necklaces.  
To begin with, I think it is fair to explain why
necklaces\cite{vilenkin_book} are produced 
in brane models, since in many papers it is discussed that ``only strings
are produced in the brane Universe''\cite{tutorial}.
Of course I think this claim is not wrong, however somewhat misleading
for non-specialists.
To explain what is misleading in the ``standard scenario'', we have a
figure in Fig.\ref{fig:fig0}. 
In general, the distance between branes may appear in the
four-dimensional effective action as a Higgs field of the effective gauge
dynamics.
At least in this case, it is natural to consider the cosmological
defects coming from the spatial variation of the Higgs field, 
which corresponds to the ``deformation'' of the
branes\cite{matsuda_deformation}. 
Is the spatial variation of the Higgs field unnatural in the brane
Universe? 
The answer is, of course, no.
One should therefore consider at least two different kinds of defects in
brane models: 
One is induced by the brane creation that is due to the spatial
variation of the tachyon condensation, while the other is induced by the
brane 
deformation that is due to the spatial variation of the brane distance.
Along the line of the above arguments, it is possible to construct
Q-ball's counterpart in brane models\cite{matsuda_Q-balls}, which can be
distinguished from the conventional Q-balls by their decay process.
We therefore have an expectation that strings can be distinguished from
the conventional ones, if one properly considers their characteristic
features.

Now let us discuss about the validity of the conventional Kibble
mechanism.
Of course the Kibble mechanism is an excellent idea that explains the
nature of the cosmological defect formation.
However, if there is oscillation of the brane distance that may be
induced by the brane inflation or by a later phase transition 
that changes the brane distance, the four-dimensional counterpart of the
brane distance (i.e. the Higgs field) oscillates in the effective
action.
In the four-dimensional counterpart, defect production induced by such
oscillation is already discussed by many 
authors, including the production of the sphaleron domain walls which
otherwise cannot be produced in the Universe\cite{matsuda_deformation}.
The defect production induced by such oscillation may or may not be
explained by the Kibble mechanism, however it should be fair to
distinguish it from the ``conventional'' Kibble mechanism.

Let us summarize the above discussion about the defect production in
the brane Universe.
Actually, it is possible to produce all kinds of defects in the brane
Universe, however it is impossible to produce defects other than the
strings {\bf simply as the result of the brane creation that is induced by
the conventional Kibble mechanism}.
One should therefore be careful about the assumption that is made in the
manuscript, which may or may not be explicit.
The necklaces are produced as the hybrid of the brane creation and the
brane deformation.
It should be noted that the stable loops of the necklaces that we will
discuss in this talk 
may appear in the four-dimensional gauge dynamics, irrespective of the
existence of the branes\cite{050906x}.
The stabilization of the necklace loops is first discussed in
Ref.\cite{matsuda_necklace} for brane models and in Ref.\cite{050906x}
for necklaces embedded in four-dimensional gauge dynamics.

In order to produce necklaces in the brane Universe, the motion in the
compactified direction is important.
I know that in the ``standard scenario'' it is sometimes discussed that
the position of the strings 
are fixed by the potential that is induced by the supersymmetry
breaking, and the position is a homogeneous parameter of the Universe
because all the decay products (Typically, they are F, D, and $(p,q)$
strings) lie (at least initially) along the same plane of the original 
hypersurface on which the tachyon condensation took place.
However, in this case one may hit upon the idea that the potential 
for a string cannot be identical to all the other kinds of the strings.
One may therefore obtain many kinds of strings that may move
independently along different
hypersurfaces, with exponentially small intersection ratios.
Moreover, I think it is not reasonable(but may be possible) to
assume that the string motion is utterly restricted by the potential
even in the most energetic epoch just after inflation.
Please remember that in general the moving (inflating) brane carries
kinetic energy, and the 
brane annihilation should be an energetic process, although one may
admit that there could be exceptional scenarios.
I therefore think that the decay products should have kinetic energy,
which is enough to climb up the potential hill at least just after
brane inflation. 

\section{PBH and DM from necklaces}
The scenario of the PBH formation from strings is initiated by
Hawking\cite{Hawking}, who utilizes the huge kinetic energy of the
shrinking loops.
However, the probability of finding loops that can
shrink into their small Schwarzschild radius is very rare due to their
random shape and motion, which weakens
the obtained bound for the string tension up to $G\mu < 10^{-6}$.
In our previous paper hep-ph/0509061\cite{0509061}, we have just extended
Hawking's idea to the networks that include monopoles attached
to the strings. 
It should be emphasized that both in the above scenarios, the kinetic
energy of the shrinking object plays crucial role.

Now let us discuss about our new idea for producing PBHs.
The most obvious discrepancy is that in our new scenario we discard the
benefit of the kinetic energy. 
We consider stable relics that are produced from necklace loops, whose
mass is large enough to turn into black holes, even after they have
dissipated their kinetic energy during the loop oscillation.
The stability of the loops is due to their windings around the
compactified space.
Of course the production of PBHs is delayed compared
to the Hawking's scenario, however
the obtained bound is much stronger than the original scenario due to
the high ($\sim 1$) production ratio.
This new mechanism of PBH production is first advocated in
hep-ph/0509062\cite{050906x}.

\begin{figure}[h]
\begin{picture}(640,330)(0,0)
\resizebox{16cm}{!}{\includegraphics{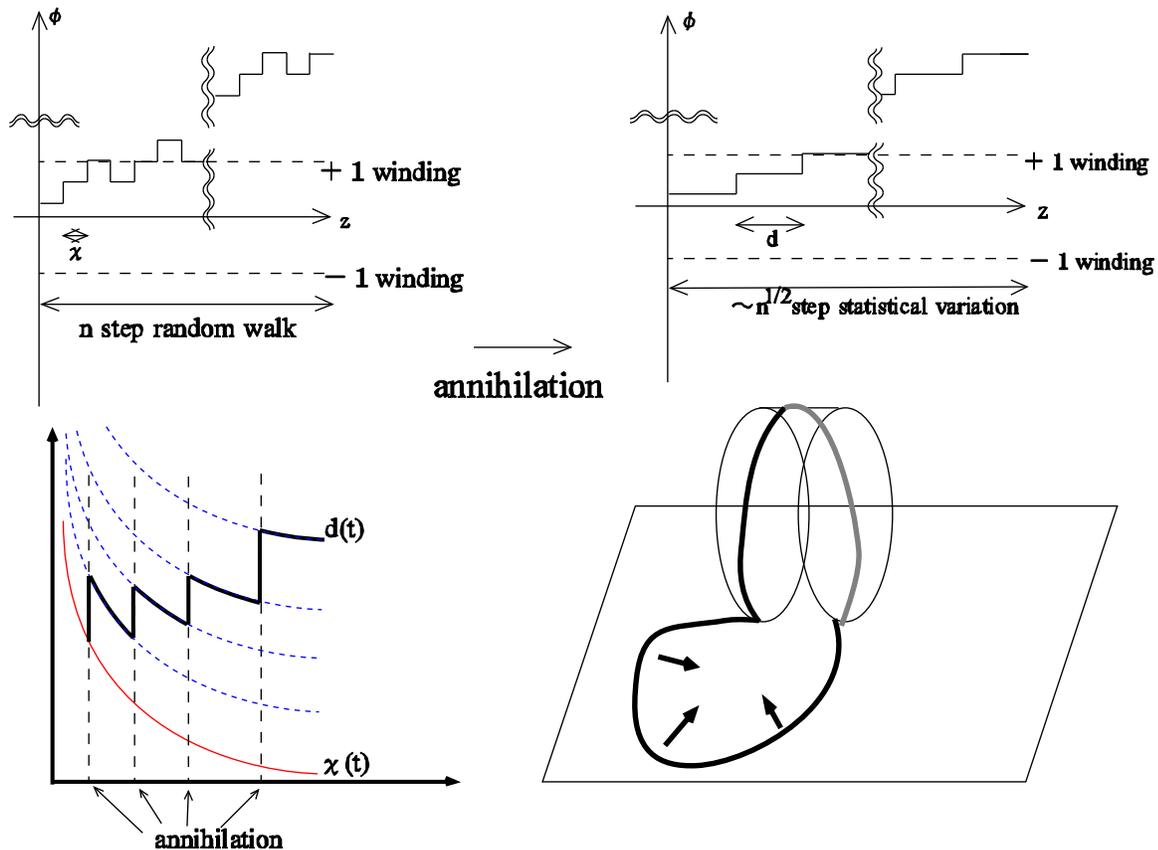}} 
\end{picture}
\caption{If the windings are stabilized by the topological reason, 
one can obtain the typical number of the windings from the simple
 statistical arguments.} 
\label{fig:count}
\end{figure}

Let us explain how one can count the winding number of a necklace
loop.
Please see Fig.\ref{fig:count}.
We introduce $\chi(t)$, which is the step length between each random
walk that corresponds to the right or the left movers in the
compactified direction. 
Since the left and the right movers can annihilate on the necklaces,
the actual distance between ``beads'' becomes much larger than
$\chi(t)$.
We therefore introduce another parameter $d(t)$, which is the typical
length between the remaining ``beads''.
Although the annihilation could be efficient, the simple statistical
argument shows that the typical number of the beads that
remain after annihilation is about $n^{1/2}$, if the initial number of
the random walk is given by $n$.
If the strings are in the scaling epoch, the typical length of the loops
is $l(t)\sim \alpha t$ when they are disconnected from the string
networks. 
Then one can obtain the typical mass of the stable relics, $M_{coils}\sim
n(t)^{1/2}m \sim [l(t)/\chi(t)]^{1/2}m$, where $m$ is the typical mass
of the beads.

Here we should note that disregarding the annihilation, $\chi(t)\propto
t^{-1}$ is already obtained in Ref.\cite{vilenkin-necklace}, which means
that $d(t)$ should evolve as $d(t)\sim t^{-1}$ at least during the short
periods between each annihilation.
Of course $d(t)$ is discontinuous at each annihilation, however the
underlying parameter $\chi(t)$ is continuous and depends on time as
$\chi(t) \sim t^{-1}$.
Using the above ideas, we can calculate the typical mass of the stable
relics that are produced from necklace loops.
The calculation of the PBH density is straightforward.
We have obtained the result 
\begin{equation}
G\mu < 10^{-21} \times 
\left[\frac{p}{10^{-2}}\right]^{4/5}
\left[\frac{\gamma}{10^{-2}}\right]^{1/5}
\left[\frac{t_n}{M_p/\mu}\right]^{3/5}
\left[\frac{d(t_n)}{M_p/\mu}\right]^{3/5}
\left[\frac{m}{10^{16} GeV}\right]^{-6/5},
\end{equation}
where we have assumed $\alpha \sim \gamma G\mu$, and $t_n$ is the time
when necklaces are produced.
$p$ is the reconnection ratio that should be $\sim 1$ for the
conventional cosmic strings, but $p \ll 1$ is possible in our case.

The string loops are produced at any time, and 
the typical mass of the stable relics depends on the time when they are
produced, because the typical length scale of
the string network increases with time both in the friction-dominated
and in the scaling epoch.
Therefore, the relics that are produced in an earlier epoch may be too
light to turn into black holes.
The ``light'' relics are the ``monopoles'' if the cosmic strings are
D-branes.
However, the ``magnetic charge'' of the ``monopoles'' may or may not be 
identical to the conventional magnetic charge of the electromagnetism.
Therefore, they are the candidate of DM, and possibly the origin of
the troublesome monopole problem only if they carry the conventional
magnetic charge. 
In our paper hep-ph/0509064\cite{050906x}, we have examined if the DM
relics can put significant bound on the tension of the cosmic strings.
We have obtained the result for $m \sim M_{GUT} \sim 10^{16}GeV$,
\begin{equation}
\label{finalresult}
G\mu< 10^{-23}\times \left[\frac{p}{10^{-2}}\right]^{9/10}
\left[\frac{1}{\beta_s}\right]^{9/10}
\left[\frac{10^{-3}}{r}\right]^{9/10}
\end{equation}
where $r$ is the mass ratio between the string part and the beads on the
necklaces, which becomes a constant in the scaling epoch\cite{050906x}.

The difficulty in lowering the typical energy scale is discussed in
Ref.\cite{matsuda_baryo} for baryogenesis and
Ref.\cite{matsuda_CMB} for the mechanism to generate density
perturbations. 

\end{document}